\documentclass[twocolumn,showpacs,preprintnumbers,amsmath,amssymb,pra]{revtex4-1}
\usepackage{hyperref}
\usepackage{graphicx}

\pdfoutput=1

\hypersetup{
	colorlinks = true,
	urlcolor   = blue,
	linkcolor  = blue,
	citecolor  = blue
}

\newcommand{\bra}[1]{\left\langle#1\right\vert}
\newcommand{\ket}[1]{\left\vert#1\right\rangle}

\begin{document}

\title{Adaptive quantum tomography}
\author{S.\,S.\,Straupe}
	\email{straups@yandex.ru}
\affiliation{Faculty of Physics, M.\,V.\,Lomonosov Moscow State University,
119001 Moscow, Russia}
\date{\today}

\begin{abstract}We provide a review of the experimental and theoretical research in the field of quantum tomography with an emphasis on recently developed adaptive protocols. Several statistical frameworks for adaptive experimental design are discussed. We argue in favor of the Bayesian approach, highlighting both its advantages for a statistical reconstruction of unknown quantum states and processes, and utility for adaptive experimental design. The discussion is supported by an analysis of several recent experimental implementations and numerical recipes.
\end{abstract}

%%% PACS numbers
%\PACS{}

\maketitle

\section{Introduction}

Quantum state tomography is a general notion standing for a set of statistical methods for the reconstruction of a density matrix, describing an unknown quantum state, using the experimental data. One can also talk about quantum process tomography, where similar techniques are applied to infer the parameters, describing an unknown transformation of the quantum state -- a quantum process. However, quantum process tomography may be reduced to quantum state tomography in a larger Hilbert space (at least on the mathematical side \cite{Choi1975,Jamiokowski1972}), so in this review we will focus on state tomography. The importance of accurate and fast procedures for statistical inference in quantum tomography is motivated by the ongoing quest of developing more and more complicated circuitry for quantum computation. The correct operation of quantum gates should be certified, and that could only be done by some kind of statistical processing of intrinsically random outcomes of quantum measurements. 

\begin{figure}[h!]
	\begin{center}
		\includegraphics[width=0.9\columnwidth]{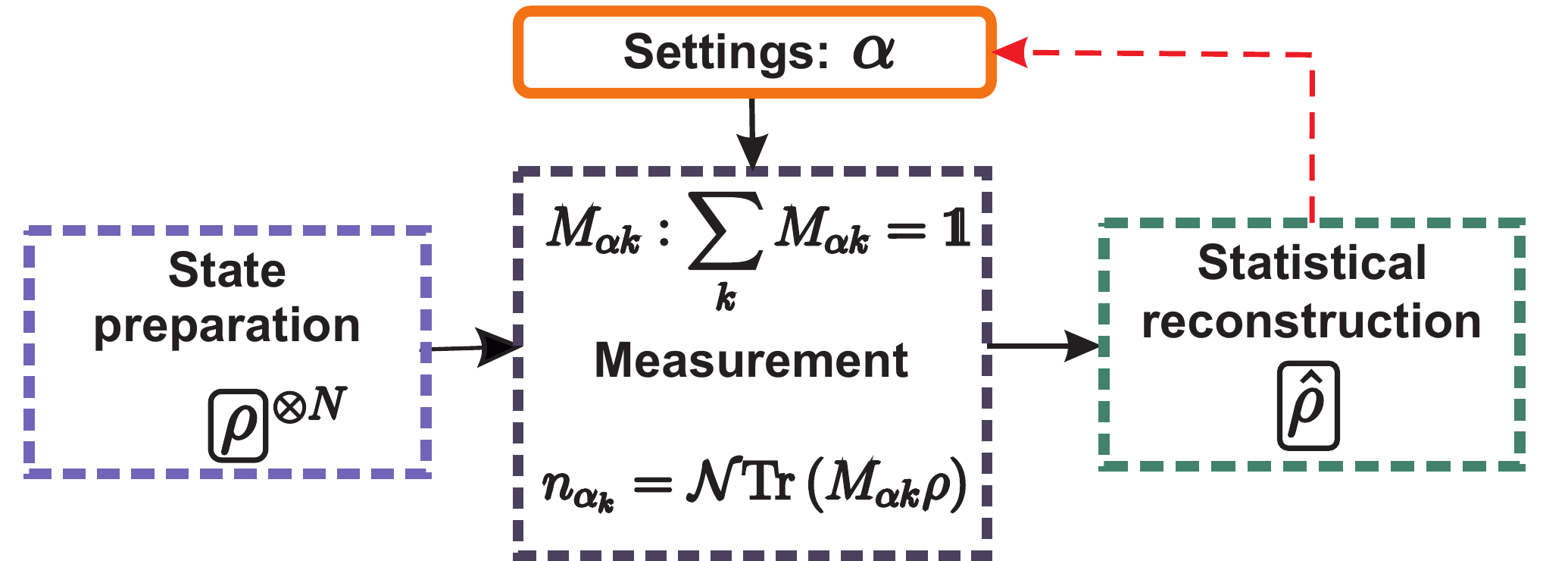}
	\end{center}
	{\caption{
			Quantum state tomography. An experimentalist has access to $N$ quantum systems in identical states $\rho$. He is free to perform measurements described by a POVM $\mathcal{M_\alpha}=\{M_{\alpha k}\}$, the choice of a particular measurement is governed by the settings of the measurement setup, which we describe by a generalized parameter $\alpha$. The outcomes of this measurements are fed to the statistical estimation procedure, which outputs an estimate $\hat{\rho}$. If the choice of measurements is performed iteratively depending on the current estimate, the procedure is called adaptive. Otherwise, the measurement settings $\alpha$ are chosen beforehand and used throughout the experiment.
		\label{Tomo_setup}}}
\end{figure}

Let us consider a generalized experiment, sketched in the Fig.~\ref{Tomo_setup}. We assume that an experimentalist has access to a large ensemble of quantum systems, prepared in identical states, described by a density matrix $\rho$. The experimentalist is free to choose a set of measurements, to be performed on the systems of interest. These measurements are in general described by a set of operators $\mathcal{M}=\{M_i\}$ corresponding to each of the outcomes, labeled by $i$, and constituting a positive operator-valued measure (POVM)\footnote{A POVM is a set of positive-semidefinite operators $\{M_i\}$, forming a resolution of identity: $\sum_i M_i = 1$}. The Born's rule of quantum theory dictates, that the probability $p_i$ of the $i$-th outcome is $p_i=\mathrm{Tr}\left(M_i\rho\right)$. In an experiment, of course, we have no access to 'true' probabilities of the outcomes. Instead we can approximate them with empirical frequencies $n_i=N_i/N$, where $N_i$ is the number of times the $i$-th outcome was observed in the series of $N$ experiments. 

Now a straightforward estimate for the unknown state would be $\hat{\rho}_{lin}$, being the solution to the set of linear equations 
\begin{equation}
	\label{linear_inversion}
	n_i=\mathrm{Tr}\left(M_i\hat{\rho}_{lin}\right).
\end{equation}  
For a unique solution to exist, the set of measurements should be tomographically complete, i.e. it should include at least $d^2$ different measurement outcomes, subject to a constraint $\sum_i p_i =1$, to determine the $d^2-1$ real parameters, describing a Hermitian density matrix with unit trace. Usually it suffices to restrict ourselves to projector-valued measures, i.e. choose $M_i$ to be one-dimensional projectors $M_i=\ket{i}\bra{i}$. Such a choice corresponds to most of the experimental situations (although we will discuss a general case later). For example, in the case of a single qubit it suffices to perform three measurements with two outcomes each, described by the following projectors:
\begin{eqnarray}
\label{6-state_POVM}
	&M_{1\pm}&=\frac{1}{2}(\ket{0}\pm i\ket{1})(\bra{0}\pm i\bra{1}),\nonumber\\
	&M_{2\pm}&=\frac{1}{2}\left(\ket{0}\pm\ket{1}\right)\left(\bra{0}\pm\bra{1}\right),\\
	&M_{3+}&=\ket{0}\bra{0},\quad M_{3-}=\ket{1}\bra{1}.\nonumber
\end{eqnarray}
The corresponding frequencies determine the, so-called, Stokes parameters: $s_i=(n_{i+}-n_{i-})/(n_{i+}+n_{i-})$, and the linear inversion estimate for the density matrix is $\hat{\rho}_{lin} = \frac{1}{2}\sum_i s_i\sigma_i$, where $\sigma_i$ are the Pauli matrices. 

The simplicity of the linear inversion tomography has a significant drawback -- since experimentally observed frequencies are random quantities, they necessarily fluctuate for every finite number of observations. That leads to solutions $\hat{\rho}_{lin}$ of (\ref{linear_inversion}) that may be unphysical, i.e. non-Hermitian, having a non-unit trace or violating a positive semi-definiteness condition. Authors of \cite{James_PRA01} empirically conclude, that it happens in about 75\% of cases for low-entropy states. More sophisticated statistical methods are required to ensure, that the reconstruction process will always produce valid physical density matrices.

\section{Maximum likelihood and other frequentist methods}

The first solution to the problem was developed by Hradil back in 1997 \cite{Hradil97}, and it was based on the well-known concept of maximum-likelihood estimation (MLE). The measured data are associated with the likelihood function, being the joint probability of observing the data (empirical frequencies $n_i$), given the particular hypothesis on what the 'true' density matrix is, considered as a function of the parameters, describing this density matrix:
\begin{equation}
	\mathcal{L}(n_i|\rho)=\prod\limits_i \mathrm{Tr}(M_i \rho).
\end{equation}
The point estimate for the 'true' density matrix in MLE is the most-likely state -- the state $\hat{\rho}_{MLE}$ maximizing the likelihood function. Now one can get rid of unphysical estimates by simply performing a constrained optimization -- restricting the space of parameters describing the density matrix to the physical manifold, for example, to the interior of a Bloch ball in the case of qubits. 

One can easily show, that if the linear inversion estimate $\hat{\rho}_{lin}$ lies inside the physical domain, it coincides with the MLE estimate \cite{BlumeKohout10}. However, if the result of the linear inversion turns out to be unphysical, the MLE procedure forces the estimate to the border of the physical domain. In this case, $\rho_{MLE}$ is usually rank-deficient, i.e. one or several of its eigenvalues are zero. Whether this situation is acceptable or not is a subject of a somewhat philosophical discussion. One of the main arguments against rank deficient estimates is that they assign essentially zero probabilities to some measurement outcomes, while that is an implausible conclusion for a finite amount of data analyzed \cite{BlumeKohout10}. There is an \emph{ad hoc} way to remedy the situation by using a so-called hedged likelihood function $\mathcal{L}_H(\rho)=\mathcal{L}(\rho)\times (\det\rho)^\beta$, with some constant $\beta$, which forces the estimate $\hat{\rho}$ to be full-rank \cite{Blume_PRL10}. Other authors, however, argue that rank deficient and even completely pure estimates are valid statistical models, as long as they describe the observed data, and one should use model selection criteria to decide, whether a full-rank or a rank deficient model should be used \cite{Bogdanov_JETP2011}. An extreme position following this line of thought leads to the \emph{model averaging} approach, where the density matrix is simultaneously reconstructed using models of different rank, and the result is averaged over the probability assigned to each model \cite{Ferrie_NJP2014a}. This approach is, however, of Bayesian nature. 

The same subtlety complicates the process of assigning 'error bars' to the point estimate. The simplest way to estimate the uncertainty in the reconstructed state is to use the Cramer-Rao bound: if the density matrix is parametrized by a vector of parameters $s_j$: $\rho = \rho(s_j)$ (for example, these can be the generalized Stokes parameters), then the variance of the estimator is bounded from below
\begin{equation}
\label{Cramer-Rao}
	(\Delta\hat{s_j})^2 \geq (I_F^{-1})_{jj},
\end{equation}
where $I_F$ is the Fisher information matrix, defined as 
\begin{equation}
\label{Fischer_matrix}
	(I_F)_{ij} = \left\langle \frac{\partial}{\partial s_i} \log\mathcal{L} \frac{\partial}{\partial s_j} \log\mathcal{L} \right\rangle.
\end{equation}

However, the Cramer-Rao bound is valid only for unbiased estimators, so it may be used only, when the estimated state is mixed and lies far enough from the boundary of the manifold of physical states. For the reasons discussed above, close to the boundary, i.e. for almost pure states, the MLE estimate tends to be biased. This region is, of course, the most interesting one, since experiments in quantum information usually aim at working with states of highest possible purity. So other, more sophisticated methods for assigning 'error bars' were proposed. For example, the concept of confidence regions was introduced in quantum state estimation by Christandl and Renner \cite{Renner_PRL12} and Blume-Kohout \cite{Blume_arXiv12}. In the latter work, confidence regions were shown to have a simple interpretation as likelihood-ratio regions -- the region $\hat{\mathcal{R}}_\alpha$ is called a likelihood-ratio region with confidence $\alpha$ if for all $\rho \in \hat{\mathcal{R}}_\alpha$, the ratio of the likelihood to its maximal value is bounded as $\mathcal{L}(\rho)/\mathcal{L}(\hat{\rho}_{MLE}) > \Lambda_\alpha$, where $\Lambda_\alpha$ is a constant, depending on the confidence level $\alpha$. Confidence regions have here the same interpretation as confidence intervals in the simple one-parametric classical statistical estimation problem: the 'true' $\rho$ is guaranteed to lie in $\hat{\mathcal{R}}_\alpha$ with probability at least $\alpha$ \cite{Renner_PRL12, Blume_arXiv12}. 

MLE is quite a computationally demanding procedure -- for high-dimensional states the dimensionality of the likelihood functional to be maximized grows exponentially. That 'dimensionality curse' can not be completely cured, however, some alternative and sometimes faster strategies exist \cite{James_PRA09}, for example one may reduce tomography to least-squares regression \cite{Vogel_PRA97,Guo_SciRep13,Padgett_PRL13}. Of course these strategies suffer from the same drawbacks, as MLE - they tend to be biased and produce rank-deficient estimates for low-entropy states.

\section{Bayesian state estimation} A different perspective on quantum state estimation is offered by Bayesian statistics \cite{BlumeKohout10,Cory_NJP16}. Here inference is based on consistent application of the Bayes rule for conditional probability \cite{Toussaint_RevModPhys11}. At first some \emph{prior} probability distribution over the state space $p(\rho)$ should be specified. The choice of this prior is the main source of controversy in Bayesian statistics in general, and we will discuss it in more details later. It is obvious, however, that the prior distribution, specified before any measurements are made, should be maximally uninformative, i.e. uniform in some sense. When the measurements are performed and the data collected, the distribution should be updated using the Bayes rule to obtain the \emph{posterior} distribution $p(\rho|\mathcal{D})\propto\mathcal{L}(\rho;\mathcal{D})p(\rho)$, where $\mathcal{D} = \{\gamma_j\}$ is the set of all measurement outcomes $\gamma_j$ -- the actual data, used for inference. This posterior distribution reflects our current knowledge about the unknown density matrix in a most complete way. A natural point estimate for the unknown state would be a \emph{Bayesian mean estimate} (BME) $\hat{\rho}_{BME}$, being a mean over the posterior distribution:
\begin{equation}
\label{BME}
	\hat{\rho}_{BME} = \int \rho p(\rho|\mathcal{D}) d\rho.
\end{equation}
BME by construction represents a physical density matrix, as long as the prior has support only on the space of physical states, which is the case for any reasonable prior. It is also full-rank for any finite set of data (unless the prior is artificially restricted to rank-deficient states), since rank-deficient states have zero measure. So in the Bayesian approach there is no problem with rank-deficient estimates. 

In contrast to MLE and other frequentist methods, besides a point estimate, Bayesian inference provides a whole distribution, which may be used to estimate error bars, as well as to obtain estimates for any properties of the state of interest as averages over the posterior. A natural way to assign 'error bars' to BME is to use the concept of \emph{credible regions} -- Bayesian counterparts of confidence intervals \cite{Englert_NJP13,Ferrie_NJP14}. An $\alpha$-credible region is the smallest set $\mathcal{X}_\alpha$ such that the probability of $\rho\in \mathcal{X}_\alpha$ is at least $1-\alpha$:
\begin{equation}
\label{Credible_region}
	\int\limits_{\mathcal{X}_\alpha} p(\rho|\mathcal{D}) d\rho \geq 1-\alpha.
\end{equation}
Precise estimation of an optimal (smallest) credible region for a given dataset under natural constraints on the density matrices set is a non-trivial problem, recently claimed to be NP-hard \cite{Gross_ArXiv16}. However for all practical purposes it can be effectively approximated. In a similar way one may determine the credible intervals for the derived properties of the state, such as purity or fidelity to some other state \cite{Englert_ArXiv16}.

We are particularly interested in the Bayesian approach to state estimation, because it allows for an easy implementation of adaptive measurement strategies. Before we discuss adaptive experimental design let us pay some attention to the practical aspects of Bayesian inference.

\section{Bayesian tomography: numerics, priors, etc.}
The main disadvantage of Bayesian methods, which until recently was preventing them from being widely used for quantum state estimation tasks, is their extreme demands for computational power. Indeed, normalization of the posterior and computation of BME in (\ref{BME}) is essentially a high-dimensional integration, which in general, is a much slower procedure, than maximizing a functional in the same space. In quantum tomography the dimensionality of the space in question grows very fast -- while for a single qubit it is a three-dimensional ball, for two-qubits it is already a manifold with 15 real dimensions. Monte Carlo algorithms are the only way to make this integration tractable \cite{Doucet_2001}. Fortunately, recently a number of fast numerical algorithms for Bayesian inference were developed, inspired by machine-learning tasks. Particularly relevant algorithms belong to the family of sequential Monte Carlo methods and were pioneered for Bayesian inference in quantum tomography by Housz\'{a}r and Houlsby \cite{Houlsby12}, and later used in the number of works \cite{Ferrie_NJP14,Cory_NJP16,Ferrie_NJP14_model,Ferrie_ArXiv16}, including experimental ones \cite{Kravtsov_PRA13,Struchalin_PRA16}.

\begin{figure}[h!]
	\includegraphics[width=0.32\columnwidth]{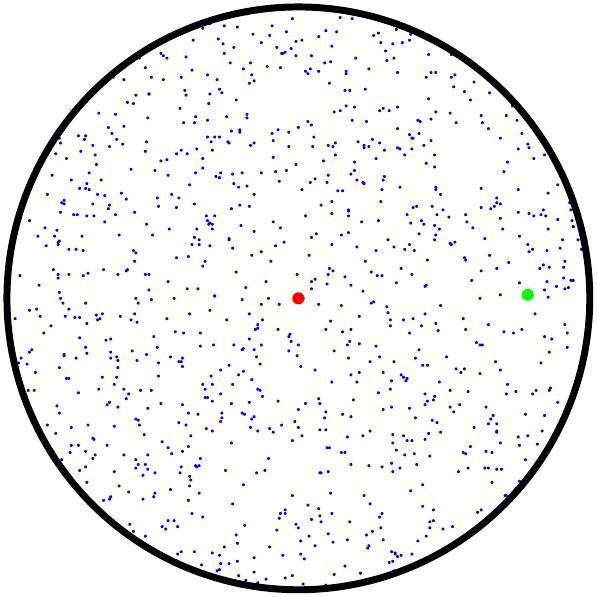}
	\includegraphics[width=0.32\columnwidth]{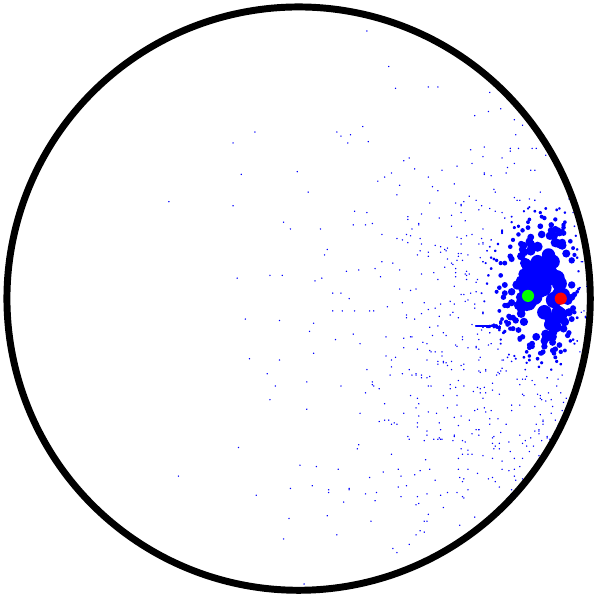}
	\includegraphics[width=0.32\columnwidth]{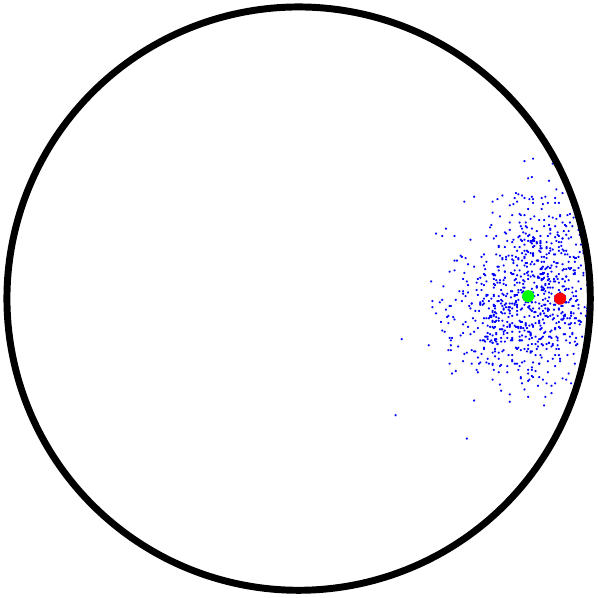}
	a)\hspace{0.3\columnwidth}b)\hspace{0.3\columnwidth}c)
	{\caption{
			Sampled distributions at different stages of the sequential Monte Carlo algorithm: (a) sampled uniform prior distribution, (b) posterior after 100 measurements, point sizes correspond to particle weights, (c) resampled posterior, weights are equalized, all information about the posterior is again contained in the spatial distribution of the particles. The green point denotes the true state, while the red one corresponds to BME according to the current posterior.}
	\label{Sampling}}
\end{figure}

The main idea behind sequential Monte Carlo is to approximate the posterior distribution by a discrete set of samples $\left\{\rho_s\right\}_{s=1}^N$ as follows\footnote{These samples are also called \emph{particles}, hence this family of algorithms is also known as \emph{particle filters}.}
\begin{equation}
\label{Particle_filter}
	p(\rho|\mathcal{D}) \approx \sum\limits_s w_s(\mathcal{D})\delta(\rho-\rho_s),
\end{equation}
where weights $w_s$ are normalized to sum to unity $\sum_s w_s =1$. At the beginning, the particles are sampled from the prior distribution, and all the weights are equalized, such that all the information about the distribution is contained in the particle positions. As new data are obtained, the weights should be updated according to the Bayes rule:
\begin{equation}
	w_s(\mathcal{D}_{j+1})= \mathbb{P}(\gamma_{j+1}|\rho_s)w_s(\mathcal{D}_j), 
\end{equation}
where $\mathbb{P}(\gamma_{j+1}|\rho_s)$ is the conditional probability to observe an outcome $\gamma_{j+1}$ under the hypothesis, that the unknown state is $\rho_s$. After every update the weights are renormalized. The main advantage here, is that the complexity of this procedure is independent of the total amount of data, since only the last term in the likelihood is used in the update. However, as the algorithm proceeds, all the weights, except a few tend to collapse to zero, reducing the quality of approximation. This situation should be avoided by adding a \emph{resampling} procedure -- periodically the particle positions $\rho_s$ should be changed and weights reset to the uniform distribution. Various methods may be used to resample the posterior correctly, such as Metropolis-Hastings sampling \cite{Hastings70} as in \cite{Houlsby12}, or a simplified Liu-West algorithm \cite{Liu2001} as in \cite{Granade12,Ferrie_ArXiv16}. We refer an interested reader to these references and to recent papers \cite{Englert_NJP15_1,Englert_NJP15_2} for a detailed discussion of Monte Carlo methods for sampling from the space of quantum states. An example of sampled distributions for the case of a rebit (a qubit restricted to the equatorial plane of the Bloch sphere) tomography is shown in Fig.~\ref{Sampling}. A sequential importance sampling algorithm with Liu-West resampling was used to generate these distributions.

When the posterior distribution is represented by weighted samples, it is straightforward to compute the Bayesian mean estimate $\hat{\rho}_{BME}=\sum_s w_s\rho_s$. Credible regions may be approximated by sorting the particles according to their weights and selecting those with the largest weights, until they sum to the desired probability $\sum w_i \geq 1-\alpha$. Then an efficient procedure exists to obtain the smallest ellipsoid enclosing this set of points, which may be reported as an estimate for an $\alpha$-credible region \cite{Ferrie_NJP14}. 

\section{How to quantify the precision of estimation?}
Statistical inference, either Bayesian or frequentist, in quantum tomography provides us with a point estimate of the 'true' state and region estimates for the 'error bars'. However, the point of tomography as a practical primitive is to diagnose the deviation of the state from some ideal one, which the experimentalist intended to prepare. For example, tomography is invaluable for debugging the quantum gates, since it is practically the only way to say, what goes wrong with the particular implementation. In order to quantify the deviation, we need to define the distance on the space of quantum states. There are numerous ways to endow the Hilbert space with geometry, we will not go into much details here and refer to an excellent book of {\.Z}yczkowski and Bengtsson \cite{Zyczkovsky_book_2006} for a thorough discussion of related mathematical subtleties. Ideally, a good notion of distance should have a statistical meaning -- it should quantify distinguishability of the two states in an experiment.

A straightforward notion of distance is provided by the \emph{Hilbert-Schmidt distance} for density matrices: $d_{HS}(\rho_1,\rho_2) = \mathrm{Tr}\left[(\rho_1-\rho_2)^2\right]$. Although convenient, the Hilbert-Schmidt distance has limited statistical significance. However, one may show, that BME is an optimal estimator in terms of the Hilbert-Schmidt distance \cite{BlumeKohout10}. The same holds for a more statistically sound notion of \emph{relative entropy} $S(\rho_1||\rho_2) = \mathrm{Tr}\left[\rho_1(\log\rho_1-\log\rho_2)\right]$, which is related to the probability of erroneously mistaking $\rho_2$ for $\rho_1$ in an experiment with a large number of copies of the state. 

The simplest distinguishability measure is the \emph{trace distance} $d_{tr}(\rho_1,\rho_2) = \frac{1}{2}\mathrm{Tr}|\rho_1-\rho_2|$. It quantifies the probability of error in distinguishing two a priori equally probable quantum states with a single measurement. As a 'single-shot' quantity it is of little relevance to quantum tomography, where one usually deals with a large number of measurements over identically prepared states. The most widely used distinguishability measure for this scenario is the \emph{fidelity}:
\begin{equation}
\label{Fidelity}
	F(\rho_1,\rho_2) = \left[\mathrm{Tr}\sqrt{\sqrt{\rho_1}\rho_2\sqrt{\rho_1}}\right]^2.
\end{equation} 
In the case of pure states it is just the transition probability $\left|\langle\psi_1|\psi_2\rangle\right|^2$. It was shown by Wootters \cite{Wootters81} that $d_{A}(\rho_1,\rho_2)=\arccos\sqrt{F(\rho_1,\rho_2)}$ is a natural quantum generalization of the statistical Fisher-Rao distance, which determines the statistical distinguishability of probability distributions in the limit of large number of samples. It is essentially the Fisher-Rao distance between the distributions of the measurement outcomes, maximized over all possible POVMs \cite{Fuchs1995}. Fidelity itself is not a proper distance, but $d_A$, called the \emph{Bures angle} is. It is convenient to use its infinitesimal version, namely the \emph{Bures distance}:
\begin{equation}
\label{Bures_distance}
	d_{B}(\rho_1,\rho_2) = \sqrt{2-2\sqrt{F(\rho_1,\rho_2)}},
\end{equation}
since it defines the same Riemannian metric on the state space and as such may not only be used to quantify the estimation quality, but also to supply an integration measure for Bayesian inference. Since $d^2_B(\rho_1,\rho_2)\approx 1-F(\rho_1,\rho_2)$ in the limit of small infidelity $1-F\ll1$, in the following we will use Bures distance and infidelity interchangeably.

Bayesian distributions should be integrated, therefore they require an integration measure on the space of quantum states. The same measure defines the notion of uniformity for an uninformative prior. One may choose an ordinary Euclidean measure and the corresponding prior, as is done in Fig.~\ref{Sampling} for illustrative purposes. However, this choice is not based on rigorous statistical grounds. We argue, that a more appropriate choice for tomographic tasks would be a Bures-uniform prior, i.e. a uniform distribution with respect to the Bures metric induced measure. Indeed, as we have seen, if one uses (in)fidelity to quantify distinguishability, the natural notion of distance is given by the Bures metric, and as every proper Riemannian metric it induces a measure in a natural way. This measure has somewhat peculiar properties, see \cite{Zyczkovsky_book_2006} for a discussion. Monte Carlo algorithms require a fast procedure to sample from the prior. Such a procedure for a Bures-uniform distribution is provided in \cite{Zyczkowski11,Mezzadri07}.

\section{Measurement sets for quantum tomography protocols}
Quantum tomography consists of two basic elements: a set of measurements performed, and an estimator, which maps the outcomes of this measurements to an estimate of the unknown state. So far we have discussed the advantages and drawbacks of various estimators, however, the performance of a tomography protocol depends as well on the proper choice of measurements. 

First of all, to obtain a unique estimate, the measurement set should be tomographically complete, i.e. the corresponding POVM should include at least $d^2$ operators. We should note, that this is not always possible to achieve in practice, especially if the state space is infinite dimensional, so methods of informationally incomplete tomography are developed to treat such cases, see \cite{Hradil_QMM13} for a review. We restrict ourselves to finite dimensional systems and informationally complete (usually over-complete) sets of measurements.

The natural question to ask is: what is the optimal measurement for a system of a given dimensionality? Let us illustrate the concept for qubits. The most widespread measurement set, corresponding to the measurement of Stokes' parameters, is given by 6 operators in (\ref{6-state_POVM}), it is obviously over-complete. The minimal measurement for a qubit should be described by a POVM with four operators. Such a POVM was introduced in \cite{Rehacek04} and consists of the following operators: $M_i=\frac{1}{4}(1+\vec{a}_i \vec{\sigma})$, where $\vec{\sigma}=\left\{\sigma_1,\sigma_2,\sigma_3\right\}$ is a vector of Pauli operators. The Bloch vectors, describing the measurements are the normal vectors to the faces of an ideal tetrahedron, with equal angles between each pair:
\begin{equation}
\label{Tetrahedron_POVM}
	\vec{a}_i\vec{a}_j=\frac{4}{3}\delta_{ij}-\frac{1}{3}.
\end{equation}
{\v{R}}eh{\'a}{\v{c}}ek et al. showed in \cite{Rehacek04}, that this POVM is an optimal one, in the sense that, for a fixed number of outcomes obtained (sample size) $N$, the variance of the Bloch vector components describing the estimated state is minimal among all possible four-element POVMs. Such a minimal measurement is quite tricky to realize experimentally, since every detector click in the experimental apparatus should be in one-to-one correspondence with one of the $M_i$, which in this case are \emph{not} the orthogonal projectors. An optical implementation for single-photon polarization qubits was demonstrated in \cite{Kurtsiefer_PRA06} with a complicated polarimeter, involving a specially designed partially-polarizing beamsplitter. 

\begin{figure}[h!]
	\begin{center}
		\includegraphics[width=0.7\columnwidth]{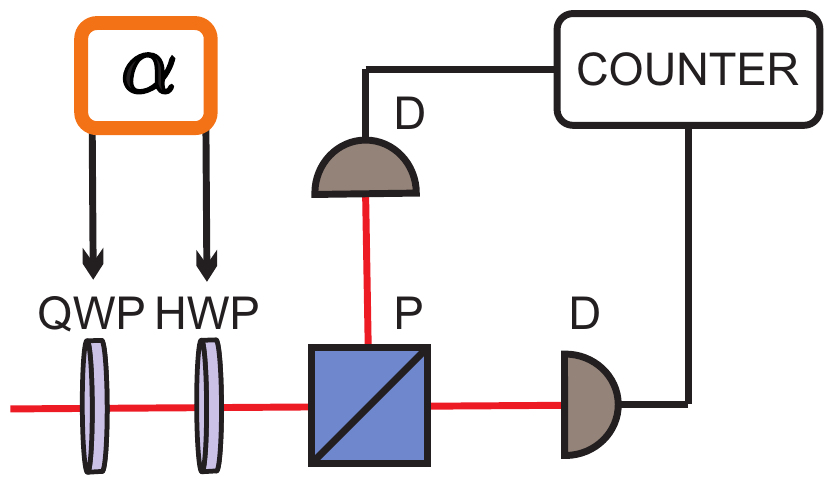}
	\end{center}
	{\caption{
			A simple setup for polarization qubit tomography. A sequence of a quarter- and half-wave plates (QWP and HWP) allows for an arbitrary unitary rotation $U_\alpha$, and the two-outcome projective measurement is performed by a polarizing beamsplitter (PBS), followed by two single-photon counting detectors (D). Such a setup realizes a POVM $M_{\alpha}=\left\{ \ket{\psi_\alpha}\bra{\psi_\alpha}, 1-\ket{\psi_\alpha}\bra{\psi_\alpha} \right\}$, where the particular measurement to be performed is specified by the waveplates angles, encoded by $\alpha$.
			\label{Polarimeter}}}
\end{figure}

In practice, fixed measurement setups, like that of \cite{Kurtsiefer_PRA06}, are rarely used. It is much easier to realize a set of two-outcome projective measurements with POVM components $M_{j\pm}=\left\{ \ket{\psi_j}\bra{\psi_j}, 1-\ket{\psi_j}\bra{\psi_j} \right\}$. For example, in a polarization qubit setting such an apparatus corresponds to a device, shown in Fig.~\ref{Polarimeter}. Measurements in this case are performed one-by-one, with different settings of the apparatus corresponding to different values of $j$. Sets of projective measurements of this kind for qubits were investigated in \cite{Jones_AoP91} and later in \cite{Langford_PRA08}. It was shown, that the Bloch vectors of optimal measurements should be isotropically arranged on the Bloch sphere. For POVMs, containing few measurements optimal sets of measurements correspond to vertices of Platonic solids, and in the limit of large number of measurements one obtains an optimal Haar-uniform POVM \cite{Jones_AoP91}. The main result of \cite{Langford_PRA08} was to show, that over-complete measurement sets, corresponding to polihedra with larger number of vertices outperform 'tetrahedral' and 'cube' (6-state) measurements in terms of fidelity. Experimental assessment of the comparative performance of the protocols based on Platonic solids was performed in \cite{Kulik_PRL10,Kulik_PRA11}, confirming the advantage of over-complete protocols with a larger number of projectors.

For higher-dimensional states, the generalizations of the 6-state POVM are POVMs based on the so-called mutually unbiased bases (MUBs) -- bases in Hilbert space, such that $\left|\langle \psi_i | \psi_j\rangle\right|^2 = \frac{1}{d}$, where $d$ is the Hilbert space dimensionality, for all states belonging to different bases. Generalizations of the tetrahedral POVM are symmetrical informationally complete measurements \cite{Caves_JMP04} (SIC-POVMs). Unfortunately, SIC-POVMs are notoriously hard to construct, and are not even proven to exist in the space of arbitrary dimensions. So most of the experimental high-dimensional tomography relies on measurements, that are simply tensor products of single-qubit ones.

\section{Performance of quantum state estimation}
Different \emph{tomography protocols} -- estimation strategies, based on a particular choice of an estimator and measurements, should be compared to each other and optimized for precision. The notion of optimality may, however, depend on the chosen figure of merit -- i.e., strictly speaking, it may be different if the estimation precision is measured in terms of (in)fidelity or, for example, Hilbert-Schmidt distance. We will therefore focus on infidelity as a figure of merit for statistical reasons discussed above. 

As usually in estimation tasks we are mostly concerned with the behavior of infidelity as a function of the \emph{sample size} $N$ -- the overall number of measurement outcomes registered. The first peculiar observation is, that infidelity behaves differently for states of different purity \cite{Bagan_PRL06,Steinberg13-journal}. Let us illustrate it by a qubit example, following \cite{Bagan_PRL06}. If qubit states are described by the Bloch vectors $\vec{s}$ and $\vec{s'}$, then the fidelity of the corresponding density matrices may be calculated as $F(\rho,\rho')= \frac{1}{2}(1+\mathbf{s}\mathbf{s}')$, where the bold letters stand for the four-dimensional vectors, defined as $\mathbf{s}=\{\sqrt{1-\left|\vec{s}\right|},\vec{s}\}$. Let $\rho$ be a pure state with a unit Bloch vector $|\vec{s}|=1$, and $\rho'$ -- a slightly mixed state with $|\vec{s'}|=1-\varepsilon$ collinear with $\vec{s}$. It is easy to calculate the infidelity: $1-F(\rho,\rho')= \varepsilon/2.$ At the same time, for mixed states which lie outside the $\varepsilon$-shell near the surface of the Bloch ball, the infidelity will be $1-F(\rho,\rho') = \varepsilon^2$. We can draw two conclusions out of this observation: 
\newline - the hard-to-estimate states lie in the vicinity of the Bloch sphere;
\newline - infidelity is hypersensitive to small eigenvalues of the density matrix.

Now consider the 6-state measurements (\ref{6-state_POVM}) and the corresponding linear inversion estimate. The estimated values of the Stokes parameters will be \cite{Steinberg13-journal}:
\begin{equation}
\label{6-state variance}
	\hat{s}_i = s_i \pm \sqrt{\frac{3}{2N}}\sqrt{1-s_i^2}.
\end{equation}
So for the length of the Bloch vector inferred from the tomographic estimate we may expect the accuracy on the order of $N^{-1/2}$, which means that the same scaling will be observed for infidelity of the states lying in the shell near the surface of the Bloch ball. The thickness of this shell is $\varepsilon\sim N^{-1/2}$. For all other states, infidelity scales as $1-F\sim N^{-1}$. 

The increasingly thin shell of hard-to-estimate states is nevertheless not to be neglected. First of all, the states of interest for most quantum information tasks are pure (to the extent, which is possible with current technology), i.e. lie in this shell. From a more statistically rigorous point of view, it is reasonable to consider the \emph{average} performance of tomographic protocols over all states. Averaging requires a prior probability distribution. A reasonable choice is again a uniform distribution with respect to the measure induced by Bures distance, since it may be argued, that such a Bures-uniform distribution is a maximally random distribution of mixed states \cite{Hall98}. The Bures-uniform prior for qubits induces the following distribution for the length of the Bloch vector $s=|\vec{s}|$ \cite{Zyczkovsky_book_2006}:
\begin{equation}
\label{Bures_radial}
	p_B(s) = \frac{4s^2}{\pi \sqrt{1-s^2}}.
\end{equation}
This distribution strongly favors the states with high purity, so the relative volume of the $\varepsilon$-shell is significant. Bagan et al. showed in \cite{Bagan2004}, that the Bures-averaged infidelity scales as $1-F\sim N^{-3/4}$. Using other sets of measurements, for example, SIC-POVMs instead of 6-state POVMs may improve the proportionality constant, but the scaling remains unaffected.

At the same time the ultimate limits of precision are known at least for qubits, and are set by collective protocols, where a complicated measurement is performed on the whole ensemble of $N$ qubits. Although unpractical, such extreme estimation strategies are useful to set the upper bounds of achievable precision for tomographic protocols. The problem was first considered for pure states of qubits in \cite{Popescu95} where the bound for fidelity, now known as the \emph{Massar-Popescu bound} was derived:
 \begin{equation}
 \label{Collective_bound_pure}
 F \leq \frac{N+1}{N+2}.
 \end{equation}    
The results for mixed states were obtained in \cite{Vidal_PRA99,Bagan2004}. It was shown \cite{Bagan2004}, that with the optimal collective measurement on $N$ qubits one can achieve the Bures-averaged fidelity of
\begin{equation}
\label{Collective_bound}
	F = 1-\left(\frac{3}{4}+\frac{4}{3\pi}\right)\frac{1}{N} + o(N^{-1}).
\end{equation}     

A natural question arises: is it possible to improve the local (measuring one qubit at a time) tomographic protocols to come closer to this ultimate bound? The answer turns out to be positive, but the structure of the protocol should be changed significantly -- it should become an adaptive estimation scheme.

\section{Adaptive quantum tomography: simplest strategies}

Standard tomography protocols described above use predetermined sets of measurements. This is not the most general local measurement scheme. Indeed one can try to benefit from using the information about the unknown state, obtained from the previous measurements, to optimize the next ones. It is evident from the analysis in the previous section, that such a strategy may help to improve the performance of tomography. 

\begin{figure}[h!]
	\begin{center}
		\includegraphics[width=0.7\columnwidth]{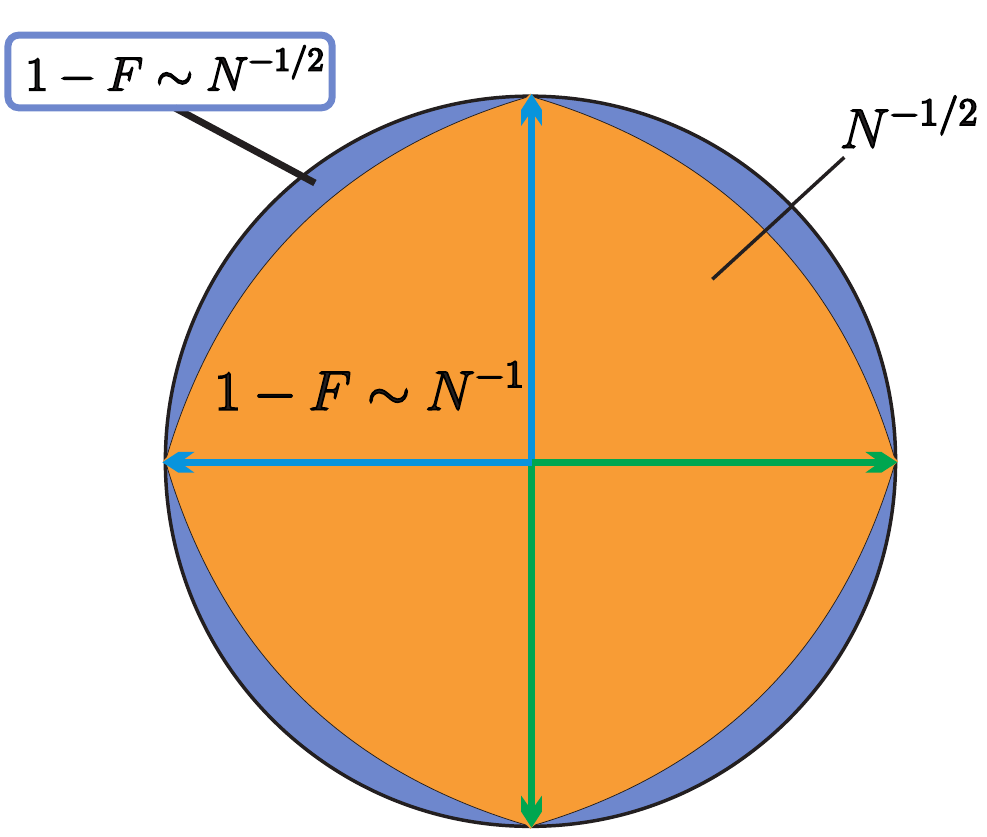}
	\end{center}
	{\caption{
		An asymptotic dependence of infidelity for a 6-state protocol on the total number of samples $N$. The scheme shows the equatorial section of the Bloch ball. The hard-to-estimate states belong to the thin layer near the surface. Note the $N^{-1}$ convergence in the directions, coinciding with the measurement basis.}
	\label{Convergence}}
\end{figure}

Note, that the variance of the estimate in (\ref{6-state variance}) depends on the 'true' state. It is minimal for states with $s_i\approx1$, which coincide with one of the basis states for the 6-state measurement in (\ref{6-state_POVM}). If the state was known a priori, the best strategy would have been to make measurements in the basis, which diagonalizes $\rho$. Now it is straightforward to come up with a simple adaptive strategy: one has to perform standard 6-state tomography on some fraction of $N_0$ qubits and obtain a preliminary estimate for $\rho$. Then one rotates the measurement basis, such that it coincides with the eigenbasis of $\rho$ and performs the remaining $N-N_0$ measurements. To the best of our knowledge, this two-step strategy was first mentioned by Gill and Massar in \cite{Massar_PRA00}, where it was suggested to attain the precision limit for local measurements on qubits $1-F = (9/4)N^{-1} + o(N^{-1})$, known as the \emph{Gill-Massar bound}. Later it was refined by \v{R}eh\'a\v{c}ek et al. \cite{Rehacek04} and Bagan et al. in \cite{Bagan_PRL06}. 

Surprisingly, the question of how to choose $N_0$ in an optimal way turns out to be not entirely trivial. Authors of \cite{Bagan_PRL06} optimize \emph{average} infidelity and find, that an asymptotically vanishing fraction of measurements $N_0=N^\alpha$  with $1/2<\alpha<1$ is enough for the rough estimate. The constant depends on a choice of prior, and for the particular case of a Bures prior $\alpha=2/3$ is found to work fine by numerical simulations. This is, however, not enough to optimize the \emph{worst-case} performance -- for almost pure states it scales as $N^{-5/6}$ \cite{Steinberg13-journal}. Optimal worst-case performance requires sacrificing $N_0=\beta N$ measurements for the first estimate. Numerical experiments in \cite{Steinberg13-journal} showed, that $\beta=1/2$ is the best choice. 

The first experimental demonstration of two-step adaptive tomography was performed for polarization qubits and confirmed the quadratic improvement in infidelity scaling \cite{Steinberg13-journal}. The experimentally obtained value for $p$ in $1-F=aN^p$ was $p=-0.90\pm0.04$. 

In a later work \cite{Guo_NPJQI16} the protocol was modified to achieve optimality in terms of the mean squared error, rather than infidelity. The main difference is that the Pauli measurements $M_{1,2}$ and $M_{3}$ (\ref{6-state_POVM}) on the second step of the protocol are performed with different probabilities: $p_1=p_2=1/(2+\sqrt{1-\hat{s}^2})$ and $p_3=\sqrt{1-\hat{s}^2}/(2+\sqrt{1-\hat{s}^2})$, where $\hat{s}$ is the length of the Bloch vector for the estimate $\hat{\rho}$, obtained at the first step, and the z-axis at the second round coincides with $\hat{\vec{s}}$. This modified protocol was experimentally shown to saturate the Gill-Massar bound for local measurements of qubits.

\section{Self-learning measurements}
A natural extension of the two-step adaptive strategy is a fully adaptive protocol, where the measurement basis is realigned with the current estimate after every measurement. Such a protocol was considered in \cite{Rehacek04} for a 4-state optimal POVM. Authors considered aligning or anti-aligning the tetrahedron (\ref{Tetrahedron_POVM}) with the current estimate for the Bloch vector $\hat{\vec{s}}$. The aligned strategy turned out to be less sensitive to misalignments with the true state, which are inevitable due to the finite estimation accuracy on any step of the protocol. The protocol was shown numerically to asymptotically reach the Massar-Popescu bound (\ref{Collective_bound_pure}), however, to the best of our knowledge, it was never implemented experimentally.

This kind of adaptive strategy is a particular case of a general approach, known as \emph{self-learning measurements} or \emph{adaptive experimental design}. It will be more instructive to look at the concept in the framework of Bayesian inference. Let $\mathbb{M}_{\alpha}$ be the POVM, corresponding to some particular choice of settings in the experimental apparatus, which we denote $\alpha$. Adaptive tomography aims at choosing $\alpha$ in an optimal way, based on the current data. Let us denote the set of observed outcomes of $n$ measurements $\mathcal{D}_n=\{\gamma_1,\ldots,\gamma_n\}$. More or less any algorithm for self-learning measurements from the Bayesian point of view may be reduced to the following scheme \cite{Toussaint_RevModPhys11}:
\newline - the first measurement setting $\alpha_0$ is chosen at random;
\newline - $n$-th measurement is chosen by optimizing a selected \emph{utility function} $U(\mathcal{\alpha,D})$ averaged over the possible measurement outcomes:
\begin{equation}
\label{Adaptivity_criterion_general}
	\alpha_{n}= \mathrm{arg}\max\limits_\alpha \sum\limits_{\gamma_{n}} p(\gamma_{n}|\alpha)U(\alpha,\mathcal{D}_{n}).
\end{equation} 
Here the probability $p(\gamma_{n}|\alpha)$ of observing a particular outcome $\gamma_{n}$ should be calculated using the current posterior distribution for the state $\rho$:
\begin{equation}
	p(\gamma_{n}|\alpha) = \int d\rho p(\rho|\mathcal{D}_{n-1}) \mathrm{Tr}\left[M_{\alpha\gamma_n}\rho\right].
\end{equation}
- the optimal measurement $\mathbb{M}_{\alpha_n}$ is performed and the optimization procedure is repeated with the new posterior $p(\rho|\mathcal{D}_n)$ to chose the next measurement.

Choosing the suitable utility function is obviously the most important issue in the whole procedure. Several possible variants are known in statistical literature. They may be roughly divided in two groups. In the first one the utility functions are constructed to optimize the parameters of the estimate. Examples are A-optimality, where the minimized quantity is the trace of the covariance matrix for the estimate, and D-optimality, where the determinant of the same matrix is minimized. The second group focuses on the \emph{information gain} of an experiment. On information-theoretic grounds the utility function here is chosen to be the expected relative entropy between the posterior $p(\rho|\mathcal{D},\alpha)$ and the prior $p(\rho)$ distributions:
\begin{equation}
\label{Information_gain}
	U(\alpha,\mathcal{D}) = \int d\rho p(\rho|\mathcal{D},\alpha)\log \frac{p(\rho|\mathcal{D},\alpha)}{p(\rho)}.
\end{equation}
Both types of utility functions were used in the design of adaptive protocols for quantum tomography. We will overview the main contributions and achievements below.

Adaptive Bayesian experimental design for optimal quantum tomography of qubits was first proposed by Fischer et al. back in 2000 \cite{Freyberger_PRA00}. They considered utility functions belonging to both classes: information gain-based, and fidelity-optimized.  Authors considered pure states only, so the fidelity-optimized utility function used was
\begin{equation}
\label{Average_fidelity_criterion}
	U(\alpha,\gamma_n) = \max\limits_{\ket{\psi}} \bra{\psi}\hat{\rho}_{n}(\gamma_n)\ket{\psi},
\end{equation}
where $\hat{\rho}_n$ is the Bayesian mean estimate over the posterior after the $n$-th measurement. Similar utility function was considered in \cite{Kalev_NJP15}. Fischer et al. numerically showed, that both strategies are advantageous over the random choice of measurements (which is the optimal non-adaptive strategy) in terms of infidelity scaling with $N$. Perhaps unsurprisingly, fidelity optimization performs slightly better, than the strategy, based on information gain. The difference was analyzed in more details in \cite{Kalev_NJP15}, where it was shown, that average fidelity maximization naturally generates MUB's for qubits in the first few iterations. Authors of \cite{Kalev_NJP15} conjectured, that this is true for any dimension, which, if true, would allow to find MUB's for the dimensions, where analytical solutions are not known. 

An optimality criterion based on the quantum Cramer-Rao inequality was put forward in \cite{Nagaoka_2005} and \cite{Fujiwara_JPA06}. Here the trace of the inverse Fischer information matrix $(I_F)^{-1}$ (\ref{Fischer_matrix}) is used as a utility function. An extension of this method leads to a quantum version of an A-optimality criterion, developed in \cite{Murao12}. We also refer to \cite{Murao12} for an overview of some other examples of useful utility functions.  

The question of which utility function to choose is a controversial one: all mentioned candidates, as well as a number of others, were successfully used in adaptive experimental design. There seems to be no consensus on the single preferable utility function, and the choice may be tailored for a particular task. In the context of quantum tomography there is, however, a generic desirable property for any utility function to be useful -- it should be efficiently computable. In the end, we want an online protocol, capable of making a decision about the best measurement at the rate of data acquisition. In this context, A-optimality, for example, taking into account all the data previously obtained, is hardly an option, unless one succeeds in finding an analytical solution for the optimization problem, which was done for qubits in \cite{Murao12}. Whether this can be done in higher dimensions is an open question. The strategy based on information gain in a measurement as a utility function looks favorable -- fast methods for its calculation were developed in the field of machine-learning and brought into the realm of quantum tomography by Husz{\'a}r and Houlsby \cite{Houlsby12}. We will discuss this and other approaches to practical adaptive tomography below.

The performance of self-learning protocols is expected to be at least not less, than that of two-step adaptive procedures, discussed in the previous paragraph. Indeed they demonstrate the optimal scaling of infidelity $1-F=aN^{-1}$, with the prefactor $a$ slightly varying depending on the particular choice of the utility function and other details of the protocol. 

\section{Implementation for qubits}

Implementation of the ideas discussed above in experiment is mostly limited by computational difficulties of solving a nonlinear optimization problem for choosing the optimal measurement. The first experimental adaptive state estimation was performed in 2001 by Hannemann et al. \cite{Wunderlich_PRA02}. The experiment was performed with qubits encoded in the hyperfine states of a single trapped $^{171}\mathrm{Yb}^{+}$ ion. Authors used the fidelity-optimized protocol of \cite{Freyberger_PRA00} with a utility function (\ref{Average_fidelity_criterion}). The optimal measurements were precomputed and stored in the look-up table, rather than be performed online. The size of the look-up table grows as $2^N$ with the number of sequential measurements $N$. That limited the number of prepare-and-measure steps in the experiment to a short sequence of $N=12$. An important experimental observation was, that the advantage of an adaptive estimation strategy over a random one is even more pronounced when the estimated state is affected by decoherence. 

Online implementation of self-learning measurements was first demonstrated by Okamoto et al. \cite{Takeuchi_PRL12}. Their work was based on the ideas of \cite{Nagaoka_2005,Fujiwara_JPA06} and actually implemented a single parameter estimation only -- the direction of polarization for linearly polarized pure photon states of the form $\ket{\psi}= \cos(\varphi)\ket{H}+\sin(\varphi)\ket{V}$, described by a single parameter $\varphi$ was reconstructed. A similar experiment for the estimation of an unknown relative phase in the entangled state of two qubits was performed in \cite{Stefanov_OpticsLett14}. Qubits were encoded in the spectral components of pulsed frequency-correlated photon pairs. Although formally the state was two-qubit, that was still a single parameter estimation experiment. 

The first experiment, where full tomography of qubits with self-learning measurements was implemented online in the course of experiment, was reported by Kravtsov et al. \cite{Kravtsov_PRA13}. In this work the optimal measurement was determined after every detection event. This was made possible by the development of a fast Bayesian protocol by Husz{\'a}r and Houlsby \cite{Houlsby12}. 

There are two main advantages of adaptive Bayesian tomography, which make it fast enough to be performed in an online manner. First of all the fast sequential importance sampling algorithm is employed, with a relatively slow Metropolis-Hastings sampling only required on the infrequent resampling stages. The second key advantage was the fast procedure for the computation of the  information gain (\ref{Information_gain}). Let us first note, that the expression for the optimal measurement at the $n$-th step of the protocol (\ref{Adaptivity_criterion_general}) with the utility function (\ref{Information_gain}) may be rewritten as 
\begin{equation}
\label{Information_gain_2}
	\alpha_n=\mathrm{arg}\max\limits_{\alpha} \left( H[p(\rho|\mathcal{D})] - \mathbb{E}_{p(\gamma_n|\alpha,\mathcal{D})} \left[ H[p(\rho|\alpha,\gamma_n,\mathcal{D})] \right] \right),
\end{equation}
where $H(p)$ is the Shanon entropy of the corresponding probability distribution, $\mathbb{E}_p(\cdot)$ denotes the expectation with respect to the probability distribution $p$, and $\mathcal{D}=\{\gamma_1,\ldots,\gamma_{n-1}\}$ is the data, obtained in the previous measurements \cite{Houlsby12}. Now the meaning of this criterion becomes intuitively clear -- it favors measurements, the outcomes of which reduce the entropy of the posterior the most. It seems, that one needs to evaluate the full posterior for every possible outcome of every measurement to compute $\alpha_n$ from (\ref{Information_gain_2}). However this is not the case, since (\ref{Information_gain_2}) may be equivalently rewritten as 
\begin{equation}
\label{Information_gain_3}
	\alpha_n= \mathrm{arg}\max\limits_{\alpha} \left( H[p(\gamma_n|\alpha,\mathcal{D})] - \mathbb{E}_{p(\rho|\mathcal{D})} \left[ H[p(\gamma_n|\alpha,\rho)] \right] \right),
\end{equation}
where $p(\gamma_n|\alpha,\mathcal{D}) = \int d\rho p(\gamma_n|\alpha,\rho)p(\rho|\mathcal{D})$ is the average predictive probability of the outcome $\gamma_n$ under the current posterior. All quantities in (\ref{Information_gain_3}) are discrete entropies and may be computed using current posterior distribution $p(\rho|\mathcal{D})$ only, represented as a set of samples (\ref{Particle_filter}). We should note, that such a reformulation is only possible for the particular utility function -- the information gain, which makes it a natural choice for a fast adaptive protocol. 

\begin{figure}[h!]
	\begin{center}
		\includegraphics[width=\columnwidth]{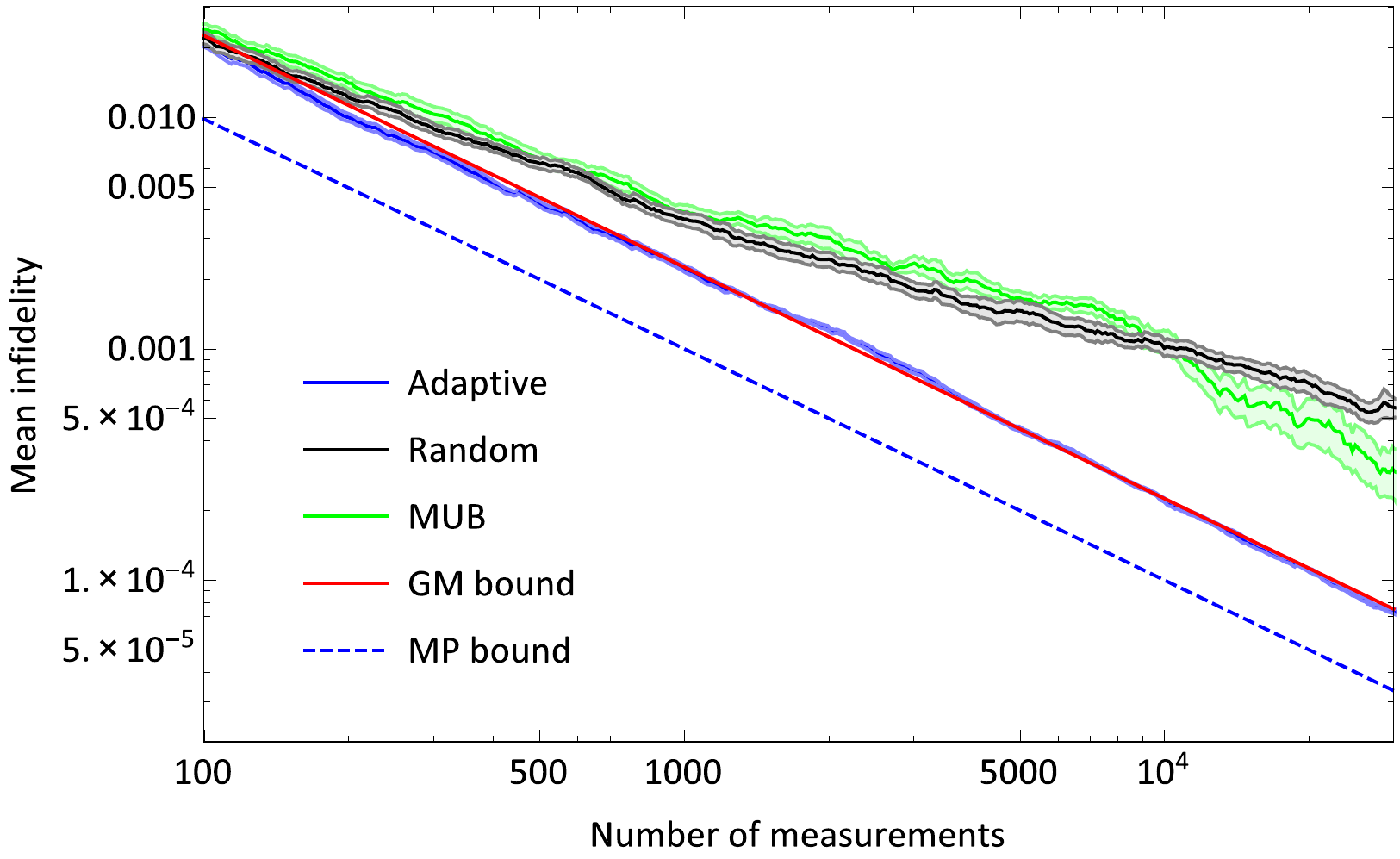}
	\end{center}
	{\caption{
			The experimentally measured dependence of infidelity on the number of measurements for the state tomography of single qubits. Solid blue line -- the adaptive Bayesian protocol, solid black line -- the Bayesian protocol with random measurements, solid green line -- the 6-state MUB protocol. Theoretical bounds are shown for comparison: solid red line -- the Gill-Massar bound $\frac{9}{4}N^{-1}$, dashed blue line -- the Massar-Popescu bound $1/(N+2)$.}
		\label{Qubit_data}}
\end{figure}

The experimental results of \cite{Kravtsov_PRA13} confirmed the advantage of the adaptive protocol over both standard 6-state MUB measurements and random measurements, sampled from the optimal uniform POVM. Since the 'true' state is not known exactly in the real experiment, the infidelity was estimated as a mean Bures distance to the Bayesian mean estimate for the posterior distribution. This quantity was shown to indeed estimate the infidelity $1-F(\rho,\hat{\rho}_{BME})$ by numerical simulations \cite{Struchalin_PRA16}. Fig.~\ref{Qubit_data} shows the experimental data for all three protocols, and the theoretical bounds for collective (Massar-Popescu) and local (Gill-Massar) measurements. Note, that the adaptive Bayesian protocol saturates the Gill-Massar bound. 

\section{Multi-qubit states and the curse of dimensionality}
The task of estimating multi-qubit states, living in high dimensional Hilbert spaces is exponentially hard. Full reconstruction with a fixed set of measurements is feasible for a Hilbert space dimension as high as 36 \cite{Kwiat_PRL05}. Numerically full tomography was recently demonstrated for a 14-qubit state (Hilbert space dimensionality of $2^{14}$) \cite{Guo_NJP16}. Experimental tomography with MUBs \cite{Padgett_PRL13} and SIC-POVMs in \cite{Boyd_PRX15} for 25- and 10-dimensional systems, respectively, was reported. Implementing adaptive strategies for something that large is hard for two main reasons. First of all, the computational complexity of high-dimensional optimization required to find an optimal measurement quickly becomes overwhelming for most of the self-learning strategies. Second, and probably more important, the problem is that optimal measurements in high dimensions almost certainly turn out to be projections on entangled states, which are extremely hard to do in experiment. There may be, however, paths to partly overcome both problems, which became recently an area of active research. 

The first route, recently taken in \cite{Struchalin_PRA16} and \cite{Guo_ArXiv2015}, is based on using optimization algorithms, that are fast enough to be implemented online even for systems of several qubits. Both works report experiments with polarization states of correlated photon pairs.

Struchalin et al. \cite{Struchalin_PRA16} used the information gain criterion in the formulation of (\ref{Information_gain_3}) to find optimal measurements for two-qubit state tomography. Unconstrained optimization according to this criterion leads to entangled projectors, which were only studied numerically. However, even if the admissible set of measurements is restricted to factorized projectors, which correspond to separate measurements on each of the qubits, the adaptive strategy based on (\ref{Information_gain_3}) provides an advantage. It was shown both numerically and experimentally, that for pure and close to pure states such factorized adaptive measurements in fact perform as well as a general adaptive strategy without any restriction on measurements. The squared Bures distance scaling in numerical simulations was found to be $d_B^2(\rho,\hat{\rho})\sim N^{a}$, with $a= -0.96\pm 0.01$. Experimentally obtained performance was somewhat lower with $a=-0.83\pm0.05$, nevertheless still way beyond that for a random protocol $a_{rnd}=-0.61\pm0.04$. The reduced performance in the experiment may to some extent be attributed to instrumental errors -- the actually performed measurements are never exactly the ones, predicted by the algorithm. The influence of the technical noise was studied and the adaptive protocol was shown to outperform random measurements even in the presence of a strong technical noise. 

\begin{figure}[h!]
	\begin{center}
		\includegraphics[width=\columnwidth]{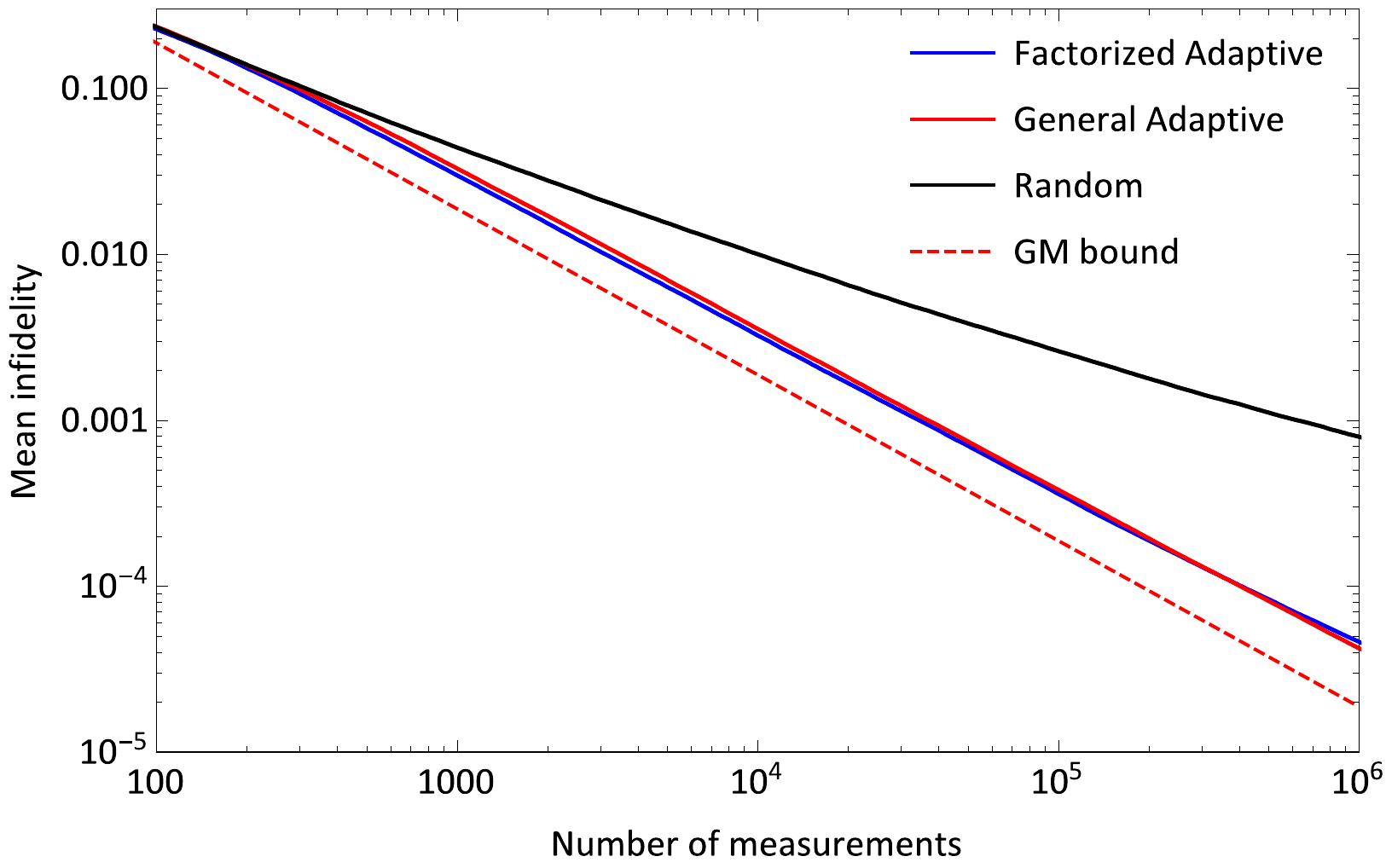}
	\end{center}
	{\caption{
			 Performance of tomography protocols for two-qubit states averaged over 1000 random pure states drawn from the Haar-uniform distribution. Solid blue line -- the adaptive Bayesian protocol with unconstrained measurements, solid red line -- the adaptive Bayesian protocol constrained to factorized measurements, solid black line -- the Bayesian protocol with random measurements. Theoretical bounds are shown for comparison: dashed red line -- the Gill-Massar bound $\frac{75}{4}N^{-1}$.
		\label{Ququart_data}}}
\end{figure}

Interestingly, the advantage of factorized adaptive measurements is pronounced for the states of high purity only, and vanishes if averaged over random states with a Bures-uniform distribution \cite{Struchalin_PRA16}. To achieve an advantage in performance on average, one has to use general non-factorized projective measurements. However, in practice one hardly deals with states randomly sampled from the Bures-uniform distribution. Usually, the states of interest have high purity. The numerical simulations in \cite{Struchalin_PRA16} showed, that if the purity of the estimated state exceeds 0.94, there is no practical advantage to use general measurements, since factorized ones perform similarly well. The average dependence of infidelity for pure states is shown in Fig.~\ref{Ququart_data}.

The approach of Qi et al. \cite{Guo_ArXiv2015} was based on state estimation by recursive linear regression used previously by the same group in \cite{Guo_SciRep13}. Optimal POVMs are chosen to minimize the mean squared error matrix for the estimate at the current step of the recursive procedure. Again two types of measurements were considered as admissible: product measurements and general unconstrained measurements. In contrast to \cite{Struchalin_PRA16}, here the adaptive protocol with unconstrained measurements was found to outperform the constrained one in terms of the prefactor on average for pure states. 

%Both protocols are numerically shown to beat the Gill-Massar bound $1-F=(75/4)N^{-1}$ for some states, while the unconstrained one does that on average over pure states as well. 

There are no experimental demonstrations going beyond two qubits so far, and that may be challenging in terms of computational time even for the fastest algorithms. There is an interesting alternative route to measurements optimization, which may turn out to be beneficial for higher-dimensional tomography. The so-called \emph{self-guided} quantum tomography was first proposed by Ferrie in \cite{Ferrie_PRL14}. The idea of the method is to treat the statistical reconstruction of a quantum state as a direct optimization problem. Considering the average infidelity $f(\hat{\rho})=\langle 1-F(\rho,\hat{\rho}) \rangle$ as a directly measurable quantity, the task is to find $\hat{\rho}$, which minimizes $f(\rho)$.  The procedure uses a stochastic algorithm, known as simultaneous perturbation stochastic approximation, which resembles the gradient descent along random directions. Starting from the random initial state $\ket{\psi_0}$ the algorithm generates measurements of the form $M_{n,\pm} = \ket{\psi_{n-1}\pm \varepsilon_n\Delta_n}\bra{\psi_{n-1}\pm \varepsilon_n\Delta_n}$, where $\Delta_n$ is a random vector consisting of $\pm1$ with equal probabilities, and $\varepsilon_n$ is the parameter chosen to ensure convergence. The gradient of infidelity is estimated via
\begin{equation}
\label{Self-guided}
	g_n=\frac{f_{n+} - f_{n-}}{2\varepsilon_n}\Delta_n,
\end{equation}  
and the next estimate for the state is $\ket{\psi_{n}}=\ket{\psi_{n-1}+\alpha_n g_n}$. The procedure converges as long as $\varepsilon_n=n^{-1/3}$, and $\alpha_n=n^{-1}$. The procedure was numerically tested for the states of up to 7 qubits \cite{Ferrie_PRL14}, which is the largest reported Hilbert space dimension for adaptive state estimation. The infidelity scales with the number of iterations as $1-F\sim n^{2/3}$. Self-guided tomography was realized experimentally in \cite{Ferrie_PRL16} for single- and two-photon polarization states. In the two-photon case the measurements were restricted to factorized projectors, similarly to \cite{Struchalin_PRA16}. 

An obvious drawback of the self-guided quantum tomography is that it works for pure states only and has no built-in procedure to provide 'error bars' for an estimate. Granade et al. \cite{Ferrie_ArXiv16} proposed to use it as an adaptive heuristic for Bayesian state estimation. This proposal solves the problem by analyzing the outcomes of measurements, returned by the self-guided algorithm, with a Bayesian estimation technique, which we discussed above in details. 

Algorithms, based on heuristics, rather than precise evaluation of the utility function, with self-guided tomography being an example, may be fast enough to be used online in experiments with few qubits. It would be interesting to see such experiments performed in the nearest future. 

\section{Conclusion}

We have provided a short review of recent advances in adaptive quantum state estimation. Several approaches were discussed with an emphasis on the methods using Bayesian experimental design. A qualitative advantage in precision offered by adaptive tomography in comparison to traditional protocols make it an attractive experimental tool. Adaptive methods have recently enjoyed experimental implementations for single and two-qubit systems, and we believe that future experiments using quantum tomography should make extensive use of similar ideas. They will be especially useful in complicated experiments with large quantum systems, where the data acquisition rate may be very low, and so extracting maximal information from the limited number of detected events is of paramount importance. We can provide recent works with multi-photon states \cite{Pan_NaturePhoton12,Pan_ArXiv16} as an example of such experiments. Although full tomography of such states may be extremely computationally demanding, some techniques of restricted state estimation exist, allowing for an efficient reconstruction of sparse or highly symmetric density matrices \cite{Eisert_PRL10,Plenio_NatureComm10,Eisert_NJP12}. Supplementing these approaches with online measurement optimization may be one of the future research directions. 

The natural limitations of the format did not allow us to cover other closely related topics. Examples include tomography with unknown POVMs, where the reconstruction procedure resembles algorithms used for noisy image processing \cite{Hradil_PRL10,Mogilevtsev_PRL13,Smith_NComm14,Silberhorn_PRA14}, which were also recently shown to benefit from adaptively optimized measurements \cite{Mogilevtsev_PRA15}. Similar techniques may prove to be useful to fight the technical noise in measurements, probably using other active machine learning algorithms. Besides tomography, Bayesian experimental design was successfully implemented for learning the parameters of the Hamiltonian, governing the evolution of the quantum system \cite{Granade_NJP2012,Wilhelm_PRL14,Wilhelm_PRA16}, phase estimation \cite{Pryde_Nature07,Bouwmeester_PRL07,Sanders_PRL10}, and characterization of coherent states \cite{Wilhelm_ArXiv15}. We believe, that we will see more fruitful applications of ideas from machine learning to quantum measurements in the nearest future.

We would like to thank G.\,I.\,Struchalin, I.\,A.\,Pogorelov, K.\,S.\,Kravtsov, I.\,V.\,Radchenko, N.\,M.\,T.\,Houlsby, Y.\,I.\,Bogdanov, and S.\,P.\,Kulik for fruitful collaboration and numerous enlightening discussions about quantum tomography. We acknowledge the support from the Russian Science Foundation Grant \#~16-12-00017.

\bibliographystyle{apsrev4-1}
\bibliography{ref_base}

\end{document}